\documentclass[runningheads]{llncs}
\usepackage{graphicx}
\usepackage{amsmath,amssymb} % define this before the line numbering.
\usepackage{color}
\usepackage{xcolor}
\usepackage{bm}
\usepackage{hyperref}
\usepackage{xcolor}
\usepackage{textcomp}
\hypersetup{
    colorlinks,
    linkcolor={blue!50!black},
    citecolor={blue!50!black},
    urlcolor={blue!80!black}
}
\usepackage{booktabs}

\newcommand{\sref}[1]{Section~\ref{#1}}
\newcommand{\fref}[1]{Figure~\ref{#1}}

\newcommand{\eref}[1]{Equation~(\ref{#1})}

\newcommand\normx[1]{\left\Vert#1\right\Vert}

\def\kmsmpc{\rm km\,s^{-1}\,Mpc^{-1}}

\begin{document}
% \linenumbers
\pagestyle{headings}
\mainmatter
\def\ECCV22SubNumber{***}  % Insert your submission number here

\title{Strong Gravitational Lensing Parameter Estimation with Vision Transformer} % Replace with your title

\titlerunning{ECCV-22 submission ID 184}

\authorrunning{ECCV-22 submission ID 184}

% \institute{Paper ID \ECCV22SubNumber}

\author{Kuan-Wei Huang\inst{1, *} \index{van Author, First E.} \and
Geoff Chih-Fan Chen\inst{2, *} \and
Po-Wen Chang\inst{3} \and
Sheng-Chieh Lin \inst{4} \and
Chia-Jung Hsu\inst{5} \and
Vishal Thengane \inst{6} \and
Joshua Yao-Yu Lin \inst{7, *} 
}
\authorrunning{K.-W. Huang et al.}
\institute{Carnegie Mellon University \and
University of California, Los Angeles \and
Ohio State University \and
University of Kentucky \and
Chalmers University of Technology \and
Mohamed bin Zayed University of Artificial Intelligence \and 
University of Illinois at Urbana-Champaign\\
$*$ equal contribution \\
\email{yaoyuyl2@illinois.edu}}

\maketitle
\begin{abstract}
Quantifying the parameters and corresponding uncertainties of hundreds of strongly lensed quasar systems holds the key to resolving one of the most important scientific questions: the Hubble constant ($H_{0}$) tension. 
The commonly used Markov chain Monte Carlo (MCMC) method has been too time-consuming to achieve this goal, yet recent work has shown that convolution neural networks (CNNs) can be an alternative with seven orders of magnitude improvement in speed.
With 31,200 simulated strongly lensed quasar images, we explore the usage of Vision Transformer (ViT) for simulated strong gravitational lensing for the first time. We show that ViT could reach competitive results compared with CNNs, and is specifically good at some lensing parameters, including the most important mass-related parameters such as the center of lens $\theta_{1}$ and $\theta_{2}$, the ellipticities $e_1$ and $e_2$, and the radial power-law slope $\gamma'$. With this promising preliminary result, we believe the ViT (or attention-based) network architecture can be an important tool for strong lensing science for the next generation of surveys. The open source of our code and data is in \url{https://github.com/kuanweih/strong_lensing_vit_resnet}.
\end{abstract}

\section{Introduction}
\label{submission}

The discovery of the accelerated expansion of the Universe \cite{PerlmutterEtal99,RiessEtal98} and observations of the Cosmic Microwave Background (CMB; e.g.,~\cite{HinshawEtal13}) established the standard cosmological paradigm: the so-called $\Lambda$ cold dark matter (CDM) model, where $\Lambda$ represents a constant dark energy density. Intriguingly the recent direct 1.7\% $H_{0}$ measurements from Type Ia supernovae (SNe), calibrated by the traditional Cepheid distance ladder ($H_{0}=73.2\pm1.3~\kmsmpc$; SH0ES collaboration \cite{RiessEtal21}), show a $4.2\sigma$ tension with the Planck results ($H_{0}=67.4\pm0.5~\kmsmpc$ \cite{planck18parameter}). However, a recent measurement of $H_{0}$ from SNe Ia calibrated by the Tip of the Red Giant Branch ($H_{0}=69.8\pm0.8(\rm{stat})\pm1.7(\rm{sys})~\kmsmpc$; CCHP collaboration \cite{FreedmanEtal20}) agrees with both the Planck and SH0ES results. The spread in these results, whether due to systematic effects or not, clearly demonstrates that it is crucial to reveal unknown systematics through different methodology.

Strongly lensed quasar system  provides such a technique to constrain $H_{0}$ at low redshift that is completely independent of the traditional distance ladder approach (e.g.,~\cite{WongEtal20,TreuMarshall16,SuyuEtal18}). When a quasar is strongly lensed by a foreground galaxy, its multiple images have light curves that are offset by a well-defined time delay, which depends on the mass profile of the lens and cosmological distances to the galaxy and the quasar \cite{Refsdal64}. However the bottleneck of using strongly lensed quasar systems is the expensive cost of computational resources and man power. With commonly used Markov chain Monte Carlo (MCMC) procedure, modeling single strongly lensed quasar system requires experienced modelers with a few months effort in order to obtain robust uncertainty estimations and up to years to check the systematics (e.g.,~\cite{WongEtal17,BirrerEtal19,RusuEtal20_H0LiCOW,GChenEtal18a,GChenEtal22_J0924,ShajibEtal20_0408}). This is infeasible as $\sim 2600$ of such systems with well-measured time delays are expected to be discovered in the upcoming survey with the Large Synoptic Survey Telescope \cite{LSSTsciencebook09,OguriMarshall10}.

\begin{figure*}
 \centering
  \includegraphics[width=0.85\linewidth]{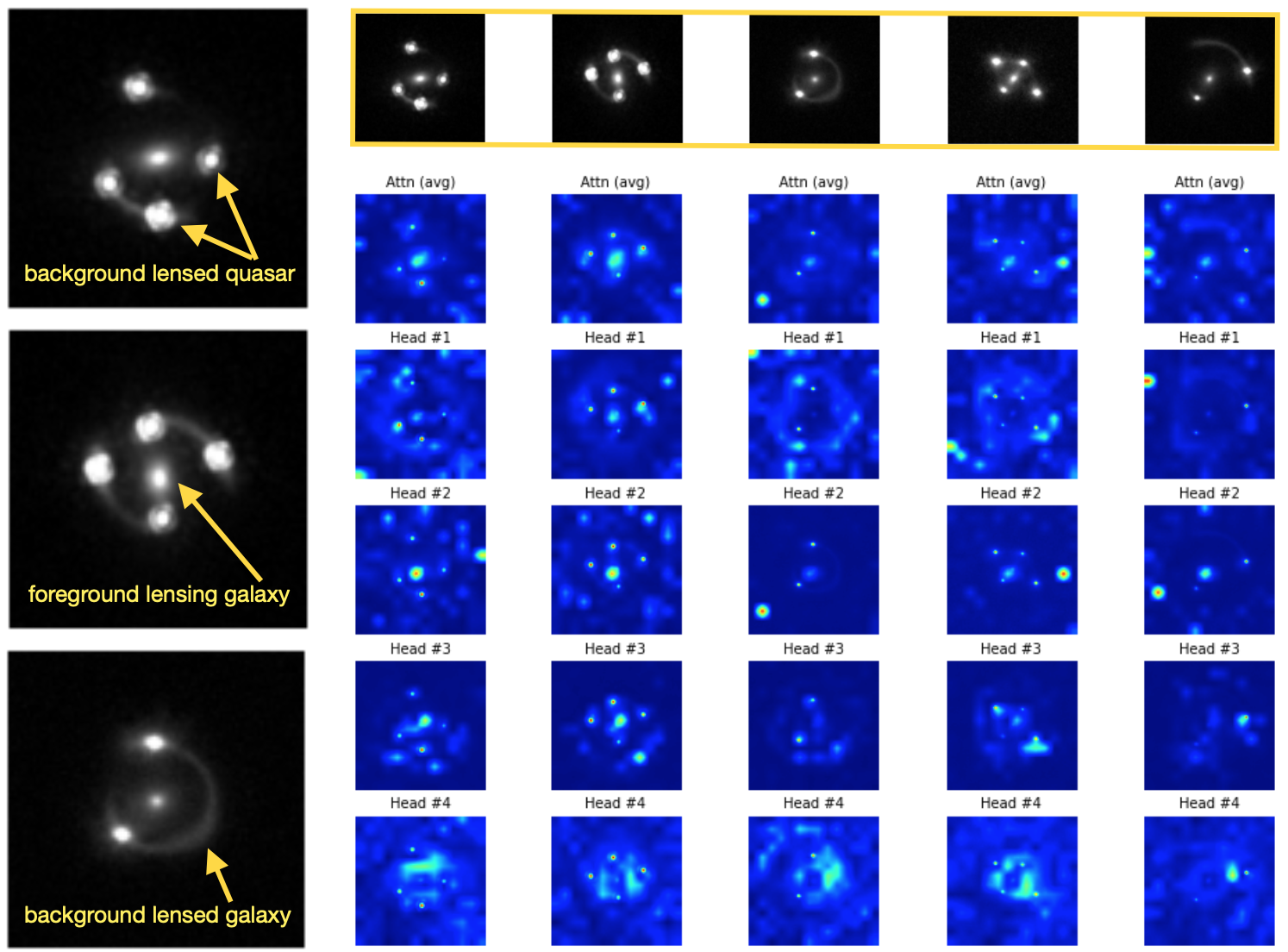}
    \caption{\textbf{Left panel}: simulated strong lensing imaging with real point spread functions (top two: space-based telescope images; bottom: ground-based adaptive-optics images). Each image contains the lensing galaxy in the middle, the multiple-lensed quasar images, and the lensed background host galaxies (arc). \textbf{Right panel}: Vision Transformer attention map: the overall average attentions are focusing on the strong lens system. Each individual head is paying attention to different subjects such as attention heads \#2 are focusing the center of lens, heads \#1 and \#3 are looking into particular lensed quasars, and heads \#4 are dealing with the arc. 
    }
  \label{fig:att_map}
\end{figure*} 

Deep learning provides a workaround for the time-consuming lens modeling task by directly mapping the underlying relationships between the input lensing images and the corresponding lensing parameters and their uncertainties. 
Hezaveh et al.~\cite{HezavehEtal17} and Perreault Levasseur et al.~\cite{PerreaultEtal17} first demonstrated that convolution neural networks (CNNs) can be an alternative to the maximum likelihood procedures with seven orders of magnitude improvement in speed. Since then, other works adpot CNN for strong lensing science related inference \cite{brehmer2019mining,wagner2021hierarchical,lin2020hunting,park2021large,Morgan:2021czo,morningstar2018analyzing,coogan2020targeted,ostdiek2022extracting,ostdiek2022image,thuruthipilly2021finding}.

In this work, instead of using traditional CNN-based models, we explore the attention-based Vision Transformer (ViT, \cite{vaswani2017attention,DosovitskiyEtal20}) that has been shown to be more robust compared with CNN-based models \cite{paul2021vision}. Furthermore, ViT retains more spatial information than ResNet \cite{raghu2021do} and hence is perfectly suitable for the strong lensing imaging as the quasar configuration and the spatially extended background lensed galaxy provide rich information on the foreground mass distribution (see \fref{fig:att_map}).

\section{Data and Models}
\label{Sec:datamodel}

In \sref{Sec:sim}, we describe the strong lensing simulation for generating the datasets in this work.
In \sref{subsec:models}, we describe the deep learning models we use to train on the simulated dataset for strong lensing parameters and uncertainty estimations.

\subsection{Simulation and Datasets}
\label{Sec:sim}

Simulating strong lensing imaging requires four major components: the mass distribution of the lensing galaxy, the source light distribution, the lens light distribution, and the point spread function (PSF), which convolves images depending on the atmosphere distortion and telescope structures. 
We use the \textsc{lenstronomy} package \cite{BirrerAmara18_lenstronomy,Birrer2021} to generate 31,200 strong lensing images with the corresponding lensing parameters for our imaging multi-regression task. 
For the mass distribution, we adapt commonly used (e.g.,~\cite{SuyuEtal13,GChenEtal22_J0924}) elliptically symmetric power-law distributions \cite{Barkana98} to model the dimensionless surface mass density of lens galaxies,
\begin{equation}
\label{eq:arclight-1}
    \kappa_{\text{pl}}
    (\theta_{1},\theta_{2})=\frac{3-
    \gamma'}{1+q}\left( 
    \frac{\theta_{\text{E}}}{\sqrt{\theta_{1}^{2}+\theta_{2}^{2}/q^2}}   \right)^{\gamma'-1},
\end{equation}
where $\gamma'$ is the radial power-law slope ($\gamma'=2$
corresponding to isothermal), $\theta_{\rm E}$ is the Einstein radius, and
$q$ is the axis ratio of the elliptical isodensity contour.
The light distribution of the lens galaxy and source galaxy are described by elliptical S$\acute{\text{e}}$rsic profile,
\begin{equation}
\label{eq:lenslight0}
    I_{\rm S}(\theta_{1},\theta_{2})=I_{\rm s}\,\mathrm{exp}\left[-k\left(\left(\frac{\sqrt{\theta_{1}^{2}+\theta_{2}^{2}/q_{\text{L}}^{2}}}{R_{\text{eff}}}\right)^{1/n_{\text{s}\acute{\text{e}}\text{rsic}}}-1\right)\right],
\end{equation}
where $I_{\rm s}$ is the amplitude, $k$ is a constant such that
$R_{\text{eff}}$ is the effective radius, $q_{\text{L}}$ is the minor-to-major axis ratio, and $n_{\text{s}\acute{\text{e}}\text{rsic}}$ is the
S$\acute{\text{e}}$rsic index \cite{Sersic68}. For the PSFs, we use six different PSF structures including three real Hubble space telescope PSFs generated by Tinytim \cite{KristHook97} and corrected by the real HST imaging \cite{GChenEtal22_J0924}, and three adaptive-optics (AO) PSFs reconstructed from ground-based Keck AO imaging \cite{GChenEtal16,GChenEtal19,GChenEtal21_AOPSF}. Three example images are shown in \fref{fig:att_map}.

We split the whole simulated dataset of 31,200 images into a training set of 27,000 images, a validation set of 3,000 images, and a test set of 1,200 images. 
We rescale each image as $3 \times 224 \times 224$ and normalize pixel values in each color channel by the mean $[0.485,\ 0.456,\ 0.406]$ and the standard deviation $[0.229,\ 0.224,\ 0.225]$ of the datasets. 
Each image has eight target variables to be predicted in this task: the Einstein radius $\theta_{\rm E}$, the ellipticities $e_1$ and $e_2$, the radial power-law slope $\gamma'$, the coordinates of mass center $\theta_{1}$ and $\theta_{2}$, the effective radius $R_{\text{eff}}$, and the S$\acute{\text{e}}$rsic index $n_{\text{s}\acute{\text{e}}\text{rsic}}$.

\subsection{Models}
\label{subsec:models}
We use the Vision Transformer (ViT) as the main model for our image multi-regression task of strong lensing parameter estimations. 
Inspired by the original Transformer models \cite{vaswani2017attention} for natural language processing tasks, Google Research proposed the ViT models \cite{DosovitskiyEtal20} for computer vision tasks. 
In this paper, we leverage the base-sized ViT model (ViT-Base), which was pre-trained on the ImageNet-21k dataset and fine-tuned on the ImageNet 2012 dataset \cite{5206848}.

Taking advantage of the transfer learning concept, we start with the pre-trained ViT-Base model downloaded from the module of \textsc{HuggingFace's Transformers} \cite{wolf-etal-2020-transformers}, and replace the last layer with a fully connected layer whose number of outputs matches the number of target variables in our regression tasks.  
The ViT model we use thus has 85,814,036 trainable parameters, patch size of 16, depth of 12, and 12 attention heads.

Alongside the ViT model, we also train a ResNet152 model \cite{HeEtal15} for the same task as a comparison between ViT and the classic benchmark CNN-based model. 
We leverage the pre-trained ResNet152 model from the \textsc{torchvision} package \cite{NEURIPS2019_9015} and modify the last layer accordingly for our multi-regression purpose.

For regression tasks, the log-likelihood can be written as a Gaussian log-likelihood \cite{Gal_Ghahramani15}. 
Thus for our task of $K$ targets, we use the negative log likelihood as the loss function \cite{PerreaultEtal17}:
\begin{equation}
\label{eq:loss_func}
\begin{split}
    {\rm Loss}_n 
        &= - \mathcal{L} \left( \bm{y}_n, \bm{\hat{y}}_n, \bm{\hat{s}}_n \right) \\
        &= \frac{1}{2} \left( \sum_{k=1}^K e^{-\hat{s}_{n, k}} \normx{y_{n, k} - \hat{y}_{n, k} }^2 + \hat{s}_{n, k} + \ln 2\pi \right)
\end{split}
\end{equation}
where ($\bm{y}_n$, $\bm{\hat{y}}_n$, $\bm{\hat{s}}_n$) are the (target, parameter estimation, uncertainty estimation) for the $n$th sample, and ($y_{n, k}$, $\hat{y}_{n, k}$, $\hat{s}_{n, k}$) are the (target, parameter estimation, uncertainty estimation) for the $n$-th sample of the $k$-th target. 
We note that in practice, working with the log-variance $\bm{\hat{s}}_n = \ln \bm{\hat{\sigma}}_n^2$ instead of the variance $\bm{\hat{\sigma}}_n^2$ improves numerical stability and avoids potential division by zero during the training process \cite{Kendall_Gal17}.
Choosing this loss function instead of the commonly used mean squared error results in the uncertainty prediction as well as the parameter prediction, which provides more statistical information than point-estimation-only predictions.

It is worth noting that we apply dropout before every hidden layers for both models with dropout rate of 0.1 to approximate Bayesian networks for the uncertainty estimate, but not for the attention layers in the ViT model. This is to include the "epistemic" uncertainties in neural networks by leaving dropout on when making predictions, together with the "aleatoric" uncertainties described by $\bm{\hat{\sigma}}_n^2$ to account for intrinsic noise from the data. We refer readers to \cite{Gal_Ghahramani15,PerreaultEtal17} for detailed discussion and derivation of the uncertainties.

Using the training set of 27,000 images, we train our ViT-Base and ResNet152 models with the loss function in \eref{eq:loss_func}, the Adam optimizer \cite{KingmaBa14} with $0.001$ for the initial learning rate, the batch size of 20. Based on the validation set of 3,000 images, we evaluate the model predictions by the mean squared error across all 8 target variables to determine the best models. We then report the performance of the best ViT and ResNet models according to the test set of 1,200 images in Section~\ref{Sec:result}.

\section{Results}
\label{Sec:result}
In this section, we present the performance of the best ViT and ResNet models on the test set of 1,200 images regards of our image multi-regression task of the strong lensing parameter and uncertainty estimation. 
Following the procedure in \cite{PerreaultEtal17}, for each model, we execute the prediction on the test set for 1000 times with dropout on to catch the epistemic uncertainty of the model.
For each parameter prediction $\bm{\hat{y}}_n$ and uncertainty prediction $\bm{\hat{\sigma}}_n$ amongst the 1000 predictions, we draw a random number from a Gaussian distribution $N\left( \bm{\hat{y}}_n, \bm{\hat{\sigma}}_n \right)$ as the prediction of the parameter. 
Therefore for each test sample, we have 1000 predicted parameters so that we take the mean and standard deviation as the final parameter and uncertainty predictions respectively.

The overall root mean square errors (RMSEs) for the lensing parameter estimation are 0.1232 for our best ViT-Base model and 0.1476 for our best ResNet152 model. The individual RMSEs for each target variable are summarized in Table~\ref{tab:param_compare}. Our models indicate that except for the Einstein radius $\theta_{\textrm{E}}$, the attention-based model ViT-Base outperforms the CNN-based model ResNet152 for all the other parameters in this image multi-regression task. Despite a higher RMSE of the Einstein radius for our best ViT than that of our best ResNet, it still reaches the benchmark precision of about 0.03 arcsec \cite{HezavehEtal17}.

\begin{table}
\caption{Comparison of RMSE of the parameter predictions between ViT and ResNet.}
\begin{center}
    \begin{tabular}{lll}
        \toprule
        Target  \makebox[2cm]{}  &  ViT \makebox[2cm]{} & ResNet \\
        \midrule
        Overall                                      & \textbf{0.1232} & 0.1476 \\
        \midrule
        $\theta_{\textrm{E}}$ [arcsec]               & 0.0302 & \textbf{0.0221} \\
        $\gamma^{\prime}$                            & \textbf{0.0789} & 0.0816 \\
        $\theta_{1}$ [arcsec]                        & \textbf{0.0033} & 0.0165 \\
        $\theta_{2}$ [arcsec]                        & \textbf{0.0036} & 0.0169 \\
        $e_{1}$                                      & \textbf{0.0278} & 0.0364 \\
        $e_{2}$                                      & \textbf{0.0206} & 0.0347 \\
        $R_{\textrm{eff}}$ [arcsec]                  & \textbf{0.0241} & 0.0487 \\
        $n_{\text{s}\acute{\text{e}}\text{rsic}}$    & \textbf{0.0790} & 0.0959 \\
        \bottomrule
    \end{tabular}
\end{center}
\label{tab:param_compare}
\end{table}

\begin{figure*}
 \centering
  \includegraphics[width=0.95\textwidth]{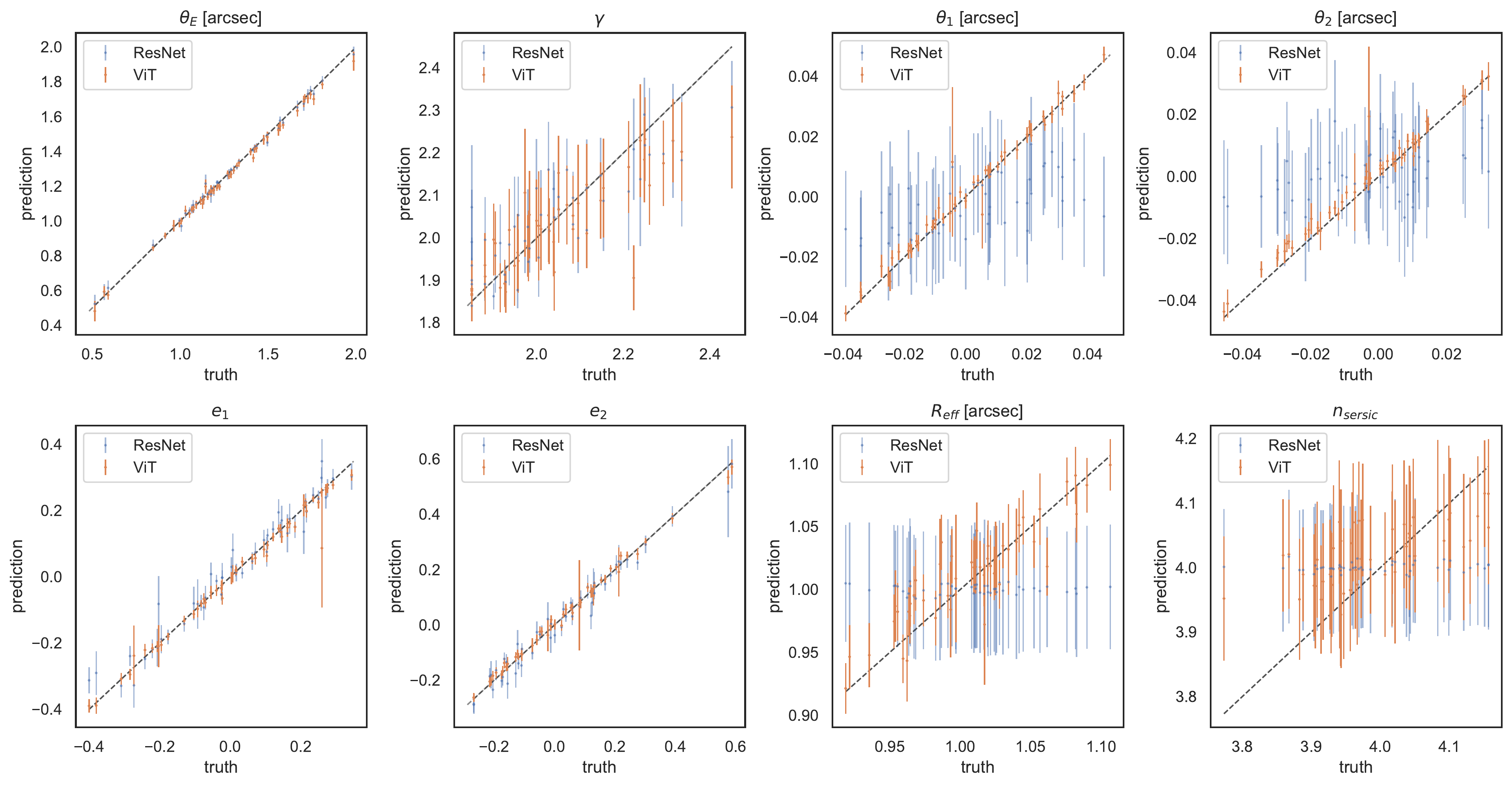}
    \caption{Comparison of predicted parameters and 1-$\sigma$ uncertainties to the ground truths for all eight targets between our best Vision Transformer (ViT) model and our best ResNet model for 50 random chosen test samples. For each target, the ground truth and the model prediction are shown on the x-axis and y-axis respectively. The ViT model outperforms the ResNet model for $\theta_1$, $\theta_2$, $e_1$, $e_2$, and $R_{\rm eff}$ while maintaining a competitive performance for the other parameters.}
  \label{fig:vit_resnet_compare}
\end{figure*}

\begin{figure*}
 \centering
  \includegraphics[width=0.95\textwidth]{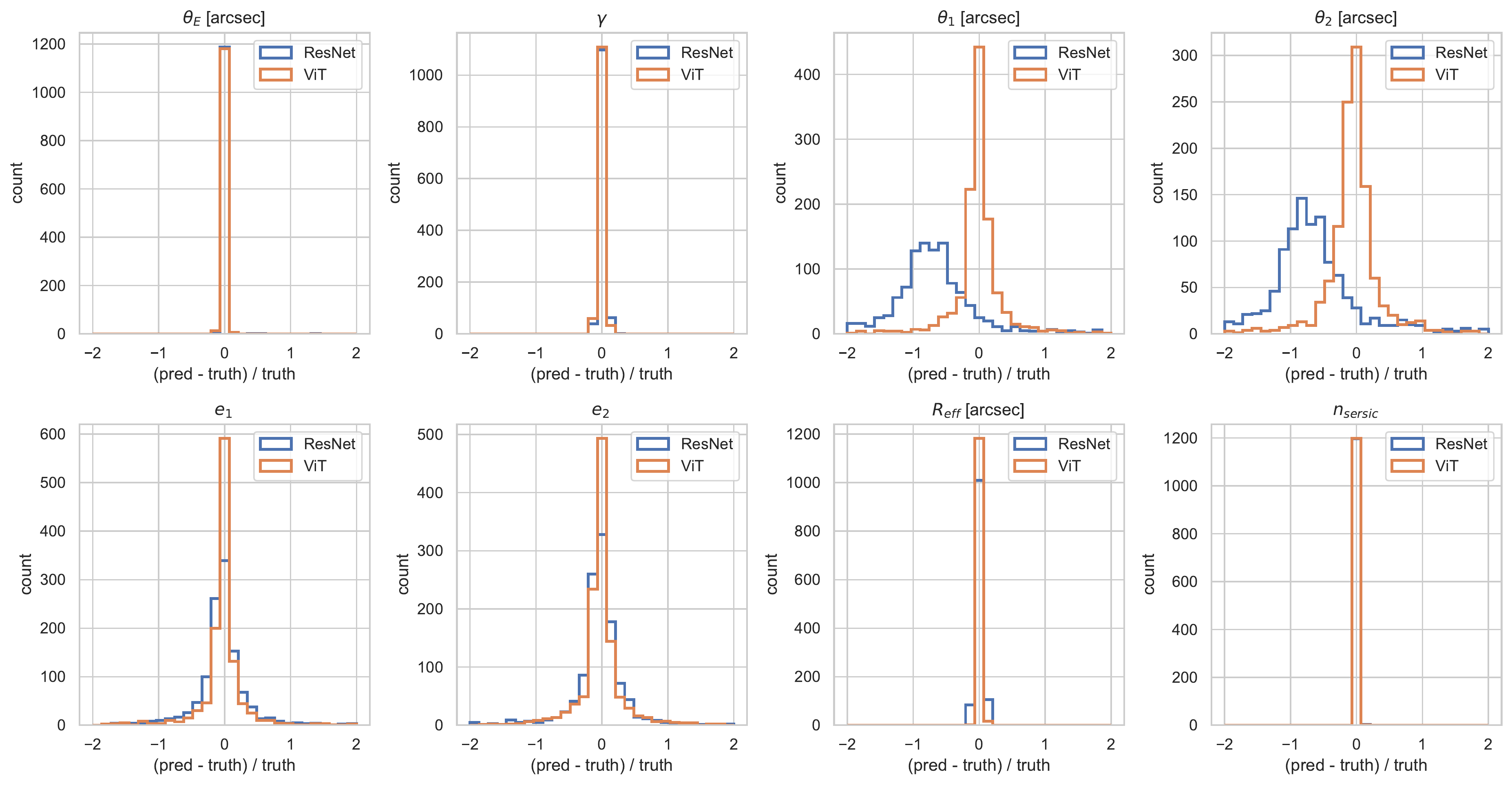}
    \caption{Comparison of predicted parameters to ground truths for all eight targets between our best Vision Transformer (ViT) model and our best ResNet model. Each panel shows the histogram of the percentage errors  between the prediction means and the ground truths of the 1200 test samples for each target. The ViT model outperforms the ResNet model for $\theta_1$, $\theta_2$, $e_1$, $e_2$, and $R_{\rm eff}$ while maintaining a competitive performance for the other parameters.}
  \label{fig:vit_resnet_compare_hist}
\end{figure*}

Using the prediction of the mean values and the corresponding uncertainties on the strong lensing parameters, we randomly select 50 test samples to illustrate the predictions from our ViT and ResNet models in \fref{fig:vit_resnet_compare}. 
Overall, the ViT model outperforms the ResNet model for $\theta_1$, $\theta_2$, $e_1$, $e_2$, and $R_{\rm eff}$ while maintaining a competitive performance for the other parameters.
We note that both models cannot well capture the features of $n_{\text{s}\acute{\text{e}}\text{rsic}}$.
In \fref{fig:vit_resnet_compare_hist}, we show the percentage error between the predictions and the ground truths for all 1200 test samples, supporting the statements above.

\section{Conclusion}
\label{Sec:conclusion}
Strongly lensed quasar systems provide unique tools to resolve the recent 4-$\sigma$ tension between the direct measurements of the $H_{0}$ and the prediction from the standard cosmological model ($\Lambda$CDM model) \cite{AbdallaEtal22,RiessEtal21,FreedmanEtal20,WongEtal20,GChenEtal19}. One of the key requirements is the mass parameter estimations of hundreds of strong lensing systems in order to achieve statistically significant results \cite{suyuEtal12b}. 
While this challenge cannot be achieved by the traditional and time-consuming MCMC method, deep neural network models can be a perfect alternative technique to efficiently achieve this goal. 
For example, Hezaveh et al.~\cite{HezavehEtal17} and Perreault Levasseur et al.~\cite{PerreaultEtal17} showed that CNN-based models could be used to estimate the values and the corresponding uncertainties of the parameters given the strong lensing images. 

In this work, we explored the recent state-of-the-art ViT as it has the advantage of capturing long-range interaction of pixels compared to CNN-based models. 
As a supervised multi-regression task, we trained ViT-Base as well as ResNet152 for the parameter and uncertainty estimations, using the dataset of 31,200 strong lensing images.

We show that ViT could reach competitive results compared with CNNs, and is specifically good at some lensing parameters, including the most important mass-related parameters such as the center of lens $\theta_{1}$ and $\theta_{2}$, the ellipticities $e_1$ and $e_2$, and the radial power-law slope $\gamma'$. With this promising preliminary result, we believe the ViT (or attention-based) network architecture can be an important tool for strong lensing science for the next generation of surveys.

%%% future work

Note that the mass distribution of real lensing galaxies are much more complicated than the simple power-law model and hence can potentially affect the $H_{0}$ measurement \cite{FalcoEtal85,GorensteinEtal88,SchneiderSluse13,XuEtal16,GomerEtal20,Kochanek20,BlumEtal20,MillonEtal20_TDCOSMOI,DingEtal21,BirrerEtal20,GChenEtal21_TDCOSMOVI}. This effect has
been illustrated with cosmological hydrodynamic simulations \cite{XuEtal16,TagoreEtal18}. In the future work, we plan to train the neural network to directly learn from realistic hydrodynamic simulations without the need of mass profile assumptions. This cannot be achieved by traditional MCMC method, while neural network is the only way to directly test this long-standing debate about the possible systematics regarding the degeneracy between lensing $H_{0}$ results and the mass 
profile assumptions. We plan to open-source our code and datasets.

\section*{Acknowledgement}
The authors thank the referees for their useful feedback and lenstronomy community for making the gravitational lensing simulation available.

\clearpage

\bibliographystyle{unsrt}
\bibliography{egbib}

\begin{thebibliography}{10}

\bibitem{PerlmutterEtal99}
S.~{Perlmutter}, G.~{Aldering}, G.~{Goldhaber}, R.~A. {Knop}, P.~{Nugent},
  P.~G. {Castro}, S.~{Deustua}, S.~{Fabbro}, and {et al.}
\newblock {Measurements of Omega and Lambda from 42 High-Redshift Supernovae}.
\newblock {\em \apj}, 517:565--586, June 1999.

\bibitem{RiessEtal98}
A.~G. {Riess}, A.~V. {Filippenko}, P.~{Challis}, A.~{Clocchiatti},
  A.~{Diercks}, P.~M. {Garnavich}, R.~L. {Gilliland}, C.~J. {Hogan}, and {et
  al.}
\newblock {Observational Evidence from Supernovae for an Accelerating Universe
  and a Cosmological Constant}.
\newblock {\em \aj}, 116:1009--1038, September 1998.

\bibitem{HinshawEtal13}
G.~{Hinshaw}, D.~{Larson}, E.~{Komatsu}, D.~N. {Spergel}, C.~L. {Bennett},
  J.~{Dunkley}, M.~R. {Nolta}, M.~{Halpern}, R.~S. {Hill}, N.~{Odegard},
  L.~{Page}, K.~M. {Smith}, J.~L. {Weiland}, B.~{Gold}, N.~{Jarosik},
  A.~{Kogut}, M.~{Limon}, S.~S. {Meyer}, G.~S. {Tucker}, E.~{Wollack}, and
  E.~L. {Wright}.
\newblock {Nine-year Wilkinson Microwave Anisotropy Probe (WMAP) Observations:
  Cosmological Parameter Results}.
\newblock {\em \apjs}, 208:19, October 2013.

\bibitem{RiessEtal21}
Adam~G. {Riess}, Stefano {Casertano}, Wenlong {Yuan}, J.~Bradley {Bowers},
  Lucas {Macri}, Joel~C. {Zinn}, and Dan {Scolnic}.
\newblock {Cosmic Distances Calibrated to 1\% Precision with Gaia EDR3
  Parallaxes and Hubble Space Telescope Photometry of 75 Milky Way Cepheids
  Confirm Tension with {\ensuremath{\Lambda}}CDM}.
\newblock {\em \apjl}, 908(1):L6, February 2021.

\bibitem{planck18parameter}
{Planck Collaboration}, N.~{Aghanim}, Y.~{Akrami}, M.~{Ashdown}, J.~{Aumont},
  C.~{Baccigalupi}, M.~{Ballardini}, A.~J. {Banday}, R.~B. {Barreiro},
  N.~{Bartolo}, S.~{Basak}, R.~{Battye}, K.~{Benabed}, J.~P. {Bernard},
  M.~{Bersanelli}, P.~{Bielewicz}, J.~J. {Bock}, J.~R. {Bond}, J.~{Borrill},
  F.~R. {Bouchet}, F.~{Boulanger}, M.~{Bucher}, C.~{Burigana}, R.~C. {Butler},
  E.~{Calabrese}, J.~F. {Cardoso}, J.~{Carron}, A.~{Challinor}, H.~C. {Chiang},
  J.~{Chluba}, L.~P.~L. {Colombo}, C.~{Combet}, D.~{Contreras}, B.~P. {Crill},
  F.~{Cuttaia}, P.~{de Bernardis}, G.~{de Zotti}, J.~{Delabrouille}, J.~M.
  {Delouis}, E.~{Di Valentino}, J.~M. {Diego}, O.~{Dor{\'e}}, M.~{Douspis},
  A.~{Ducout}, X.~{Dupac}, S.~{Dusini}, G.~{Efstathiou}, F.~{Elsner}, T.~A.
  {En{\ss}lin}, H.~K. {Eriksen}, Y.~{Fantaye}, M.~{Farhang}, J.~{Fergusson},
  R.~{Fernandez-Cobos}, F.~{Finelli}, F.~{Forastieri}, M.~{Frailis}, A.~A.
  {Fraisse}, E.~{Franceschi}, A.~{Frolov}, S.~{Galeotta}, S.~{Galli},
  K.~{Ganga}, R.~T. {G{\'e}nova-Santos}, M.~{Gerbino}, T.~{Ghosh},
  J.~{Gonz{\'a}lez-Nuevo}, K.~M. {G{\'o}rski}, S.~{Gratton}, A.~{Gruppuso},
  J.~E. {Gudmundsson}, J.~{Hamann}, W.~{Handley}, F.~K. {Hansen}, D.~{Herranz},
  S.~R. {Hildebrandt}, E.~{Hivon}, Z.~{Huang}, A.~H. {Jaffe}, W.~C. {Jones},
  A.~{Karakci}, E.~{Keih{\"a}nen}, R.~{Keskitalo}, K.~{Kiiveri}, J.~{Kim},
  T.~S. {Kisner}, L.~{Knox}, N.~{Krachmalnicoff}, M.~{Kunz}, H.~{Kurki-Suonio},
  G.~{Lagache}, J.~M. {Lamarre}, A.~{Lasenby}, M.~{Lattanzi}, C.~R. {Lawrence},
  M.~{Le Jeune}, P.~{Lemos}, J.~{Lesgourgues}, F.~{Levrier}, A.~{Lewis},
  M.~{Liguori}, P.~B. {Lilje}, M.~{Lilley}, V.~{Lindholm},
  M.~{L{\'o}pez-Caniego}, P.~M. {Lubin}, Y.~Z. {Ma}, J.~F.
  {Mac{\'\i}as-P{\'e}rez}, G.~{Maggio}, D.~{Maino}, N.~{Mandolesi},
  A.~{Mangilli}, A.~{Marcos-Caballero}, M.~{Maris}, P.~G. {Martin},
  M.~{Martinelli}, E.~{Mart{\'\i}nez-Gonz{\'a}lez}, S.~{Matarrese}, N.~{Mauri},
  J.~D. {McEwen}, P.~R. {Meinhold}, A.~{Melchiorri}, A.~{Mennella},
  M.~{Migliaccio}, M.~{Millea}, S.~{Mitra}, M.~A. {Miville-Desch{\^e}nes},
  D.~{Molinari}, L.~{Montier}, G.~{Morgante}, A.~{Moss}, P.~{Natoli}, H.~U.
  {N{\o}rgaard-Nielsen}, L.~{Pagano}, D.~{Paoletti}, B.~{Partridge},
  G.~{Patanchon}, H.~V. {Peiris}, F.~{Perrotta}, V.~{Pettorino},
  F.~{Piacentini}, L.~{Polastri}, G.~{Polenta}, J.~L. {Puget}, J.~P. {Rachen},
  M.~{Reinecke}, M.~{Remazeilles}, A.~{Renzi}, G.~{Rocha}, C.~{Rosset},
  G.~{Roudier}, J.~A. {Rubi{\~n}o-Mart{\'\i}n}, B.~{Ruiz-Granados},
  L.~{Salvati}, M.~{Sandri}, M.~{Savelainen}, D.~{Scott}, E.~P.~S. {Shellard},
  C.~{Sirignano}, G.~{Sirri}, L.~D. {Spencer}, R.~{Sunyaev}, A.~S. {Suur-Uski},
  J.~A. {Tauber}, D.~{Tavagnacco}, M.~{Tenti}, L.~{Toffolatti}, M.~{Tomasi},
  T.~{Trombetti}, L.~{Valenziano}, J.~{Valiviita}, B.~{Van Tent}, L.~{Vibert},
  P.~{Vielva}, F.~{Villa}, N.~{Vittorio}, B.~D. {Wandelt}, I.~K. {Wehus},
  M.~{White}, S.~D.~M. {White}, A.~{Zacchei}, and A.~{Zonca}.
\newblock {Planck 2018 results. VI. Cosmological parameters}.
\newblock {\em \aap}, 641:A6, September 2020.

\bibitem{FreedmanEtal20}
Wendy~L. {Freedman}, Barry~F. {Madore}, Taylor {Hoyt}, In~Sung {Jang}, Rachael
  {Beaton}, Myung~Gyoon {Lee}, Andrew {Monson}, Jill {Neeley}, and Jeffrey
  {Rich}.
\newblock {Calibration of the Tip of the Red Giant Branch}.
\newblock {\em \apj}, 891(1):57, March 2020.

\bibitem{WongEtal20}
Kenneth~C. {Wong}, Sherry~H. {Suyu}, Geoff C.~F. {Chen}, Cristian~E. {Rusu},
  Martin {Millon}, Dominique {Sluse}, Vivien {Bonvin}, Christopher~D.
  {Fassnacht}, Stefan {Taubenberger}, Matthew~W. {Auger}, Simon {Birrer}, James
  H.~H. {Chan}, Frederic {Courbin}, Stefan {Hilbert}, Olga {Tihhonova}, Tommaso
  {Treu}, Adriano {Agnello}, Xuheng {Ding}, Inh {Jee}, Eiichiro {Komatsu},
  Anowar~J. {Shajib}, Alessandro {Sonnenfeld}, Roger~D. {Blandford}, L{\'e}on
  V.~E. {Koopmans}, Philip~J. {Marshall}, and Georges {Meylan}.
\newblock {H0LiCOW {\textendash} XIII. A 2.4 per cent measurement of H$_{0}$
  from lensed quasars: 5.3{\ensuremath{\sigma}} tension between early- and
  late-Universe probes}.
\newblock {\em \mnras}, 498(1):1420--1439, October 2020.

\bibitem{TreuMarshall16}
T.~{Treu} and P.~J. {Marshall}.
\newblock {Time delay cosmography}.
\newblock {\em \aapr}, 24:11, July 2016.

\bibitem{SuyuEtal18}
Sherry~H. {Suyu}, Tzu-Ching {Chang}, Fr{\'e}d{\'e}ric {Courbin}, and Teppei
  {Okumura}.
\newblock {Cosmological Distance Indicators}.
\newblock {\em \ssr}, 214(5):91, Aug 2018.

\bibitem{Refsdal64}
S.~{Refsdal}.
\newblock {On the possibility of determining Hubble's parameter and the masses
  of galaxies from the gravitational lens effect}.
\newblock {\em \mnras}, 128:307, 1964.

\bibitem{WongEtal17}
K.~C. {Wong}, S.~H. {Suyu}, M.~W. {Auger}, V.~{Bonvin}, F.~{Courbin}, C.~D.
  {Fassnacht}, A.~{Halkola}, C.~E. {Rusu}, D.~{Sluse}, A.~{Sonnenfeld},
  T.~{Treu}, T.~E. {Collett}, S.~{Hilbert}, L.~V.~E. {Koopmans}, P.~J.
  {Marshall}, and N.~{Rumbaugh}.
\newblock {H0LiCOW - IV. Lens mass model of HE 0435-1223 and blind measurement
  of its time-delay distance for cosmology}.
\newblock {\em \mnras}, 465:4895--4913, March 2017.

\bibitem{BirrerEtal19}
S.~{Birrer}, T.~{Treu}, C.~E. {Rusu}, V.~{Bonvin}, C.~D. {Fassnacht}, J.~H.~H.
  {Chan}, A.~{Agnello}, A.~J. {Shajib}, G.~C.-F. {Chen}, M.~{Auger},
  F.~{Courbin}, S.~{Hilbert}, D.~{Sluse}, S.~H. {Suyu}, K.~C. {Wong},
  P.~{Marshall}, B.~C. {Lemaux}, and G.~{Meylan}.
\newblock {H0LiCOW - IX. Cosmographic analysis of the doubly imaged quasar SDSS
  1206+4332 and a new measurement of the Hubble constant}.
\newblock {\em \mnras}, 484:4726--4753, April 2019.

\bibitem{RusuEtal20_H0LiCOW}
Cristian~E. {Rusu}, Kenneth~C. {Wong}, Vivien {Bonvin}, Dominique {Sluse},
  Sherry~H. {Suyu}, Christopher~D. {Fassnacht}, James H.~H. {Chan}, Stefan
  {Hilbert}, Matthew~W. {Auger}, Alessandro {Sonnenfeld}, Simon {Birrer},
  Frederic {Courbin}, Tommaso {Treu}, Geoff C.~F. {Chen}, Aleksi {Halkola},
  L{\'e}on V.~E. {Koopmans}, Philip~J. {Marshall}, and Anowar~J. {Shajib}.
\newblock {H0LiCOW XII. Lens mass model of WFI2033-4723 and blind measurement
  of its time-delay distance and H$_{0}$}.
\newblock {\em \mnras}, 498(1):1440--1468, October 2020.

\bibitem{GChenEtal18a}
Geoff C.-F. {Chen}, James H.~H. {Chan}, Vivien {Bonvin}, Christopher~D.
  {Fassnacht}, Karina {Rojas}, Martin {Millon}, Fred {Courbin}, Sherry~H.
  {Suyu}, Kenneth~C. {Wong}, Dominique {Sluse}, Tommaso {Treu}, Anowar~J.
  {Shajib}, Jen-Wei {Hsueh}, David~J. {Lagattuta}, L{\'e}on V.~E. {Koopmans},
  Simona {Vegetti}, and John~P. {McKean}.
\newblock {Constraining the microlensing effect on time delays with a new
  time-delay prediction model in H$_{0}$ measurements}.
\newblock {\em \mnras}, 481(1):1115--1125, Nov 2018.

\bibitem{GChenEtal22_J0924}
Geoff C.~F. {Chen}, Christopher~D. {Fassnacht}, Sherry~H. {Suyu}, L{\'e}on
  V.~E. {Koopmans}, David~J. {Lagattuta}, John~P. {McKean}, Matt~W. {Auger},
  Simona {Vegetti}, and Tommaso {Treu}.
\newblock {SHARP - VIII. J 0924+0219 lens mass distribution and time-delay
  prediction through adaptive-optics imaging}.
\newblock {\em \mnras}, April 2022.

\bibitem{ShajibEtal20_0408}
A.~J. {Shajib}, S.~{Birrer}, T.~{Treu}, A.~{Agnello}, E.~J. {Buckley-Geer},
  J.~H.~H. {Chan}, L.~{Christensen}, C.~{Lemon}, H.~{Lin}, M.~{Millon},
  J.~{Poh}, C.~E. {Rusu}, D.~{Sluse}, C.~{Spiniello}, G.~C.~F. {Chen},
  T.~{Collett}, F.~{Courbin}, C.~D. {Fassnacht}, J.~{Frieman}, A.~{Galan},
  D.~{Gilman}, A.~{More}, T.~{Anguita}, M.~W. {Auger}, V.~{Bonvin},
  R.~{McMahon}, G.~{Meylan}, K.~C. {Wong}, T.~M.~C. {Abbott}, J.~{Annis},
  S.~{Avila}, K.~{Bechtol}, D.~{Brooks}, D.~{Brout}, D.~L. {Burke}, A.~{Carnero
  Rosell}, M.~{Carrasco Kind}, J.~{Carretero}, F.~J. {Castander},
  M.~{Costanzi}, L.~N. {da Costa}, J.~{De Vicente}, S.~{Desai}, J.~P.
  {Dietrich}, P.~{Doel}, A.~{Drlica-Wagner}, A.~E. {Evrard}, D.~A. {Finley},
  B.~{Flaugher}, P.~{Fosalba}, J.~{Garc{\'\i}a-Bellido}, D.~W. {Gerdes},
  D.~{Gruen}, R.~A. {Gruendl}, J.~{Gschwend}, G.~{Gutierrez}, D.~L.
  {Hollowood}, K.~{Honscheid}, D.~{Huterer}, D.~J. {James}, T.~{Jeltema},
  E.~{Krause}, N.~{Kuropatkin}, T.~S. {Li}, M.~{Lima}, N.~{MacCrann}, M.~A.~G.
  {Maia}, J.~L. {Marshall}, P.~{Melchior}, R.~{Miquel}, R.~L.~C. {Ogando},
  A.~{Palmese}, F.~{Paz-Chinch{\'o}n}, A.~A. {Plazas}, A.~K. {Romer},
  A.~{Roodman}, M.~{Sako}, E.~{Sanchez}, B.~{Santiago}, V.~{Scarpine},
  M.~{Schubnell}, D.~{Scolnic}, S.~{Serrano}, I.~{Sevilla-Noarbe}, M.~{Smith},
  M.~{Soares-Santos}, E.~{Suchyta}, G.~{Tarle}, D.~{Thomas}, A.~R. {Walker},
  and Y.~{Zhang}.
\newblock {STRIDES: a 3.9 per cent measurement of the Hubble constant from the
  strong lens system DES J0408-5354}.
\newblock {\em \mnras}, 494(4):6072--6102, March 2020.

\bibitem{LSSTsciencebook09}
{LSST Science Collaboration}, Paul~A. {Abell}, Julius {Allison}, Scott~F.
  {Anderson}, John~R. {Andrew}, J.~Roger~P. {Angel}, Lee {Armus}, David
  {Arnett}, S.~J. {Asztalos}, Tim~S. {Axelrod}, Stephen {Bailey}, D.~R.
  {Ballantyne}, Justin~R. {Bankert}, Wayne~A. {Barkhouse}, Jeffrey~D. {Barr},
  L.~Felipe {Barrientos}, Aaron~J. {Barth}, James~G. {Bartlett}, Andrew~C.
  {Becker}, Jacek {Becla}, Timothy~C. {Beers}, Joseph~P. {Bernstein}, Rahul
  {Biswas}, Michael~R. {Blanton}, Joshua~S. {Bloom}, John~J. {Bochanski}, Pat
  {Boeshaar}, Kirk~D. {Borne}, Marusa {Bradac}, W.~N. {Brandt}, Carrie~R.
  {Bridge}, Michael~E. {Brown}, Robert~J. {Brunner}, James~S. {Bullock},
  Adam~J. {Burgasser}, James~H. {Burge}, David~L. {Burke}, Phillip~A.
  {Cargile}, Srinivasan {Chand rasekharan}, George {Chartas}, Steven~R.
  {Chesley}, You-Hua {Chu}, David {Cinabro}, Mark~W. {Claire}, Charles~F.
  {Claver}, Douglas {Clowe}, A.~J. {Connolly}, Kem~H. {Cook}, Jeff {Cooke},
  Asantha {Cooray}, Kevin~R. {Covey}, Christopher~S. {Culliton}, Roelof {de
  Jong}, Willem~H. {de Vries}, Victor~P. {Debattista}, Francisco {Delgado},
  Ian~P. {Dell'Antonio}, Saurav {Dhital}, Rosanne {Di Stefano}, Mark
  {Dickinson}, Benjamin {Dilday}, S.~G. {Djorgovski}, Gregory {Dobler}, Ciro
  {Donalek}, Gregory {Dubois-Felsmann}, Josef {Durech}, Ardis {Eliasdottir},
  Michael {Eracleous}, Laurent {Eyer}, Emilio~E. {Falco}, Xiaohui {Fan},
  Christopher~D. {Fassnacht}, Harry~C. {Ferguson}, Yanga~R. {Fernandez},
  Brian~D. {Fields}, Douglas {Finkbeiner}, Eduardo~E. {Figueroa}, Derek~B.
  {Fox}, Harold {Francke}, James~S. {Frank}, Josh {Frieman}, Sebastien
  {Fromenteau}, Muhammad {Furqan}, Gaspar {Galaz}, A.~{Gal-Yam}, Peter
  {Garnavich}, Eric {Gawiser}, John {Geary}, Perry {Gee}, Robert~R. {Gibson},
  Kirk {Gilmore}, Emily~A. {Grace}, Richard~F. {Green}, William~J. {Gressler},
  Carl~J. {Grillmair}, Salman {Habib}, J.~S. {Haggerty}, Mario {Hamuy}, Alan~W.
  {Harris}, Suzanne~L. {Hawley}, Alan~F. {Heavens}, Leslie {Hebb}, Todd~J.
  {Henry}, Edward {Hileman}, Eric~J. {Hilton}, Keri {Hoadley}, J.~B. {Holberg},
  Matt~J. {Holman}, Steve~B. {Howell}, Leopoldo {Infante}, Zeljko {Ivezic},
  Suzanne~H. {Jacoby}, Bhuvnesh {Jain}, {R}, {Jedicke}, M.~James {Jee},
  J.~{Garrett Jernigan}, Saurabh~W. {Jha}, Kathryn~V. {Johnston}, R.~Lynne
  {Jones}, Mario {Juric}, Mikko {Kaasalainen}, {Styliani}, {Kafka}, Steven~M.
  {Kahn}, Nathan~A. {Kaib}, Jason {Kalirai}, Jeff {Kantor}, Mansi~M.
  {Kasliwal}, Charles~R. {Keeton}, Richard {Kessler}, Zoran {Knezevic}, Adam
  {Kowalski}, Victor~L. {Krabbendam}, K.~Simon {Krughoff}, Shrinivas
  {Kulkarni}, Stephen {Kuhlman}, Mark {Lacy}, Sebastien {Lepine}, Ming {Liang},
  Amy {Lien}, Paulina {Lira}, Knox~S. {Long}, Suzanne {Lorenz}, Jennifer~M.
  {Lotz}, R.~H. {Lupton}, Julie {Lutz}, Lucas~M. {Macri}, Ashish~A. {Mahabal},
  Rachel {Mandelbaum}, Phil {Marshall}, Morgan {May}, Peregrine~M. {McGehee},
  Brian~T. {Meadows}, Alan {Meert}, Andrea {Milani}, Christopher~J. {Miller},
  Michelle {Miller}, David {Mills}, Dante {Minniti}, David {Monet}, Anjum~S.
  {Mukadam}, Ehud {Nakar}, Douglas~R. {Neill}, Jeffrey~A. {Newman}, Sergei
  {Nikolaev}, Martin {Nordby}, Paul {O'Connor}, Masamune {Oguri}, John
  {Oliver}, Scot~S. {Olivier}, Julia~K. {Olsen}, Knut {Olsen}, Edward~W.
  {Olszewski}, Hakeem {Oluseyi}, Nelson~D. {Padilla}, Alex {Parker}, Joshua
  {Pepper}, John~R. {Peterson}, Catherine {Petry}, Philip~A. {Pinto}, James~L.
  {Pizagno}, Bogdan {Popescu}, Andrej {Prsa}, Veljko {Radcka}, M.~Jordan
  {Raddick}, Andrew {Rasmussen}, Arne {Rau}, Jeonghee {Rho}, James~E. {Rhoads},
  Gordon~T. {Richards}, Stephen~T. {Ridgway}, Brant~E. {Robertson}, Rok
  {Roskar}, Abhijit {Saha}, Ata {Sarajedini}, Evan {Scannapieco}, Terry
  {Schalk}, Rafe {Schindler}, Samuel {Schmidt}, Sarah {Schmidt}, Donald~P.
  {Schneider}, German {Schumacher}, Ryan {Scranton}, Jacques {Sebag}, Lynn~G.
  {Seppala}, Ohad {Shemmer}, Joshua~D. {Simon}, M.~{Sivertz}, Howard~A.
  {Smith}, J.~{Allyn Smith}, Nathan {Smith}, Anna~H. {Spitz}, Adam {Stanford},
  Keivan~G. {Stassun}, Jay {Strader}, Michael~A. {Strauss}, Christopher~W.
  {Stubbs}, Donald~W. {Sweeney}, Alex {Szalay}, Paula {Szkody}, Masahiro
  {Takada}, Paul {Thorman}, David~E. {Trilling}, Virginia {Trimble}, Anthony
  {Tyson}, Richard {Van Berg}, Daniel {Vand en Berk}, Jake {VanderPlas}, Licia
  {Verde}, Bojan {Vrsnak}, Lucianne~M. {Walkowicz}, Benjamin~D. {Wand elt},
  Sheng {Wang}, Yun {Wang}, Michael {Warner}, Risa~H. {Wechsler}, Andrew~A.
  {West}, Oliver {Wiecha}, Benjamin~F. {Williams}, Beth {Willman}, David
  {Wittman}, Sidney~C. {Wolff}, W.~Michael {Wood-Vasey}, Przemek {Wozniak},
  Patrick {Young}, Andrew {Zentner}, and Hu~{Zhan}.
\newblock {LSST Science Book, Version 2.0}.
\newblock {\em arXiv e-prints}, page arXiv:0912.0201, Dec 2009.

\bibitem{OguriMarshall10}
M.~{Oguri} and P.~J. {Marshall}.
\newblock {Gravitationally lensed quasars and supernovae in future wide-field
  optical imaging surveys}.
\newblock {\em \mnras}, 405:2579--2593, July 2010.

\bibitem{HezavehEtal17}
Yashar~D. {Hezaveh}, Laurence {Perreault Levasseur}, and Philip~J. {Marshall}.
\newblock {Fast automated analysis of strong gravitational lenses with
  convolutional neural networks}.
\newblock {\em \nat}, 548(7669):555--557, August 2017.

\bibitem{PerreaultEtal17}
Laurence {Perreault Levasseur}, Yashar~D. {Hezaveh}, and Risa~H. {Wechsler}.
\newblock {Uncertainties in Parameters Estimated with Neural Networks:
  Application to Strong Gravitational Lensing}.
\newblock {\em \apjl}, 850(1):L7, November 2017.

\bibitem{brehmer2019mining}
Johann Brehmer, Siddharth Mishra-Sharma, Joeri Hermans, Gilles Louppe, and Kyle
  Cranmer.
\newblock Mining for dark matter substructure: Inferring subhalo population
  properties from strong lenses with machine learning.
\newblock {\em The Astrophysical Journal}, 886(1):49, 2019.

\bibitem{wagner2021hierarchical}
Sebastian Wagner-Carena, Ji~Won Park, Simon Birrer, Philip~J Marshall, Aaron
  Roodman, Risa~H Wechsler, LSST Dark Energy~Science Collaboration, et~al.
\newblock Hierarchical inference with bayesian neural networks: An application
  to strong gravitational lensing.
\newblock {\em The Astrophysical Journal}, 909(2):187, 2021.

\bibitem{lin2020hunting}
Joshua Yao-Yu Lin, Hang Yu, Warren Morningstar, Jian Peng, and Gilbert Holder.
\newblock {Hunting for Dark Matter Subhalos in Strong Gravitational Lensing
  with Neural Networks}.
\newblock In {\em {34th Conference on Neural Information Processing Systems}},
  10 2020.

\bibitem{park2021large}
Ji~Won Park, Sebastian Wagner-Carena, Simon Birrer, Philip~J Marshall, Joshua
  Yao-Yu Lin, Aaron Roodman, LSST Dark Energy~Science Collaboration, et~al.
\newblock Large-scale gravitational lens modeling with bayesian neural networks
  for accurate and precise inference of the hubble constant.
\newblock {\em The Astrophysical Journal}, 910(1):39, 2021.

\bibitem{Morgan:2021czo}
Robert Morgan, Brian Nord, Simon Birrer, Joshua Yao-Yu Lin, and Jason Poh.
\newblock {deeplenstronomy: A dataset simulation package for strong
  gravitational lensing}.
\newblock {\em J. Open Source Softw.}, 6(58):2854, 2021.

\bibitem{morningstar2018analyzing}
Warren~R Morningstar, Yashar~D Hezaveh, Laurence~Perreault Levasseur, Roger~D
  Blandford, Philip~J Marshall, Patrick Putzky, and Risa~H Wechsler.
\newblock Analyzing interferometric observations of strong gravitational lenses
  with recurrent and convolutional neural networks.
\newblock {\em arXiv preprint arXiv:1808.00011}, 2018.

\bibitem{coogan2020targeted}
Adam Coogan, Konstantin Karchev, and Christoph Weniger.
\newblock Targeted likelihood-free inference of dark matter substructure in
  strongly-lensed galaxies.
\newblock In {\em 34th Conference on Neural Information Processing Systems}, 10
  2020.

\bibitem{ostdiek2022extracting}
Bryan Ostdiek, Ana Diaz~Rivero, and Cora Dvorkin.
\newblock Extracting the subhalo mass function from strong lens images with
  image segmentation.
\newblock {\em The Astrophysical Journal}, 927(1), 3 2022.

\bibitem{ostdiek2022image}
Bryan Ostdiek, Ana Diaz~Rivero, and Cora Dvorkin.
\newblock {Image segmentation for analyzing galaxy-galaxy strong lensing
  systems}.
\newblock {\em Astron. Astrophys.}, 657:L14, 2022.

\bibitem{thuruthipilly2021finding}
Hareesh Thuruthipilly, Adam Zadrozny, and Agnieszka Pollo.
\newblock Finding strong gravitational lenses through self-attention.
\newblock {\em arXiv preprint arXiv:2110.09202}, 2021.

\bibitem{vaswani2017attention}
Ashish Vaswani, Noam Shazeer, Niki Parmar, Jakob Uszkoreit, Llion Jones,
  Aidan~N Gomez, {\L}ukasz Kaiser, and Illia Polosukhin.
\newblock Attention is all you need.
\newblock {\em Advances in neural information processing systems}, 30, 2017.

\bibitem{DosovitskiyEtal20}
Alexey Dosovitskiy, Lucas Beyer, Alexander Kolesnikov, Dirk Weissenborn,
  Xiaohua Zhai, Thomas Unterthiner, Mostafa Dehghani, Matthias Minderer, Georg
  Heigold, Sylvain Gelly, Jakob Uszkoreit, and Neil Houlsby.
\newblock An image is worth 16x16 words: Transformers for image recognition at
  scale.
\newblock In {\em 9th International Conference on Learning Representations,
  {ICLR} 2021, Virtual Event, Austria, May 3-7, 2021}. OpenReview.net, 2021.

\bibitem{paul2021vision}
Sayak Paul and Pin-Yu Chen.
\newblock Vision transformers are robust learners.
\newblock In {\em AAAI}, 2022.

\bibitem{raghu2021do}
Maithra Raghu, Thomas Unterthiner, Simon Kornblith, Chiyuan Zhang, and Alexey
  Dosovitskiy.
\newblock Do vision transformers see like convolutional neural networks?
\newblock In A.~Beygelzimer, Y.~Dauphin, P.~Liang, and J.~Wortman Vaughan,
  editors, {\em Advances in Neural Information Processing Systems}, 2021.

\bibitem{BirrerAmara18_lenstronomy}
Simon {Birrer} and Adam {Amara}.
\newblock {lenstronomy: Multi-purpose gravitational lens modelling software
  package}.
\newblock {\em Physics of the Dark Universe}, 22:189--201, Dec 2018.

\bibitem{Birrer2021}
Simon Birrer, Anowar~J. Shajib, Daniel Gilman, Aymeric Galan, Jelle Aalbers,
  Martin Millon, Robert Morgan, Giulia Pagano, Ji~Won Park, Luca Teodori,
  Nicolas Tessore, Madison Ueland, Lyne~Van de~Vyvere, Sebastian Wagner-Carena,
  Ewoud Wempe, Lilan Yang, Xuheng Ding, Thomas Schmidt, Dominique Sluse, Ming
  Zhang, and Adam Amara.
\newblock lenstronomy ii: A gravitational lensing software ecosystem.
\newblock {\em Journal of Open Source Software}, 6(62):3283, 2021.

\bibitem{SuyuEtal13}
S.~H. {Suyu}, M.~W. {Auger}, S.~{Hilbert}, P.~J. {Marshall}, M.~{Tewes},
  T.~{Treu}, C.~D. {Fassnacht}, L.~V.~E. {Koopmans}, D.~{Sluse}, R.~D.
  {Blandford}, F.~{Courbin}, and G.~{Meylan}.
\newblock {Two Accurate Time-delay Distances from Strong Lensing: Implications
  for Cosmology}.
\newblock {\em \apj}, 766:70, April 2013.

\bibitem{Barkana98}
R.~{Barkana}.
\newblock {Fast Calculation of a Family of Elliptical Mass Gravitational Lens
  Models}.
\newblock {\em \apj}, 502:531, August 1998.

\bibitem{Sersic68}
J.~L. {S{\'e}rsic}.
\newblock {\em {Atlas de galaxias australes}}.
\newblock Cordoba, Argentina: Observatorio Astronomico, 1968, 1968.

\bibitem{KristHook97}
J.~E. {Krist} and R.~N. {Hook}.
\newblock {NICMOS PSF variations and tiny tim simulations}.
\newblock In S.~{Casertano}, R.~{Jedrzejewski}, T.~{Keyes}, and M.~{Stevens},
  editors, {\em The 1997 HST Calibration Workshop with a New Generation of
  Instruments, p. 192}, page 192, January 1997.

\bibitem{GChenEtal16}
G.~C.-F. {Chen}, S.~H. {Suyu}, K.~C. {Wong}, C.~D. {Fassnacht}, T.~{Chiueh},
  A.~{Halkola}, I.~S. {Hu}, M.~W. {Auger}, L.~V.~E. {Koopmans}, D.~J.
  {Lagattuta}, J.~P. {McKean}, and S.~{Vegetti}.
\newblock {SHARP - III. First use of adaptive-optics imaging to constrain
  cosmology with gravitational lens time delays}.
\newblock {\em \mnras}, 462:3457--3475, November 2016.

\bibitem{GChenEtal19}
Geoff C.-F. {Chen}, Christopher~D. {Fassnacht}, Sherry~H. {Suyu}, Cristian~E.
  {Rusu}, James H.~H. {Chan}, Kenneth~C. {Wong}, Matthew~W. {Auger}, Stefan
  {Hilbert}, Vivien {Bonvin}, Simon {Birrer}, Martin {Millon}, L{\'e}on V.~E.
  {Koopmans}, David~J. {Lagattuta}, John~P. {McKean}, Simona {Vegetti},
  Frederic {Courbin}, Xuheng {Ding}, Aleksi {Halkola}, Inh {Jee}, Anowar~J.
  {Shajib}, Dominique {Sluse}, Alessandro {Sonnenfeld}, and Tommaso {Treu}.
\newblock {A SHARP view of H0LiCOW: H$_{0}$ from three time-delay gravitational
  lens systems with adaptive optics imaging}.
\newblock {\em \mnras}, 490(2):1743--1773, December 2019.

\bibitem{GChenEtal21_AOPSF}
Geoff C.-F. {Chen}, Tommaso {Treu}, Christopher~D. {Fassnacht}, Sam {Ragland},
  Thomas {Schmidt}, and Sherry~H. {Suyu}.
\newblock {Point spread function reconstruction of adaptive-optics imaging:
  meeting the astrometric requirements for time-delay cosmography}.
\newblock {\em \mnras}, 508(1):755--761, November 2021.

\bibitem{5206848}
Jia Deng, Wei Dong, Richard Socher, Li-Jia Li, Kai Li, and Li~Fei-Fei.
\newblock Imagenet: A large-scale hierarchical image database.
\newblock In {\em 2009 IEEE Conference on Computer Vision and Pattern
  Recognition}, pages 248--255, 2009.

\bibitem{wolf-etal-2020-transformers}
Thomas Wolf, Lysandre Debut, Victor Sanh, Julien Chaumond, Clement Delangue,
  Anthony Moi, Pierric Cistac, Tim Rault, Rémi Louf, Morgan Funtowicz, Joe
  Davison, Sam Shleifer, Patrick von Platen, Clara Ma, Yacine Jernite, Julien
  Plu, Canwen Xu, Teven~Le Scao, Sylvain Gugger, Mariama Drame, Quentin Lhoest,
  and Alexander~M. Rush.
\newblock Transformers: State-of-the-art natural language processing.
\newblock In {\em Proceedings of the 2020 Conference on Empirical Methods in
  Natural Language Processing: System Demonstrations}, pages 38--45, Online,
  October 2020. Association for Computational Linguistics.

\bibitem{HeEtal15}
Kaiming {He}, Xiangyu {Zhang}, Shaoqing {Ren}, and Jian {Sun}.
\newblock {Deep Residual Learning for Image Recognition}.
\newblock {\em arXiv e-prints}, page arXiv:1512.03385, December 2015.

\bibitem{NEURIPS2019_9015}
Adam Paszke, Sam Gross, Francisco Massa, Adam Lerer, James Bradbury, Gregory
  Chanan, Trevor Killeen, Zeming Lin, Natalia Gimelshein, Luca Antiga, Alban
  Desmaison, Andreas Kopf, Edward Yang, Zachary DeVito, Martin Raison, Alykhan
  Tejani, Sasank Chilamkurthy, Benoit Steiner, Lu~Fang, Junjie Bai, and Soumith
  Chintala.
\newblock Pytorch: An imperative style, high-performance deep learning library.
\newblock In H.~Wallach, H.~Larochelle, A.~Beygelzimer, F.~d\textquotesingle
  Alch\'{e}-Buc, E.~Fox, and R.~Garnett, editors, {\em Advances in Neural
  Information Processing Systems 32}, pages 8024--8035. Curran Associates,
  Inc., 2019.

\bibitem{Gal_Ghahramani15}
Yarin Gal and Zoubin Ghahramani.
\newblock Dropout as a bayesian approximation: Representing model uncertainty
  in deep learning.
\newblock In Maria~Florina Balcan and Kilian~Q. Weinberger, editors, {\em
  Proceedings of The 33rd International Conference on Machine Learning},
  volume~48 of {\em Proceedings of Machine Learning Research}, pages
  1050--1059, New York, New York, USA, 20--22 Jun 2016. PMLR.

\bibitem{Kendall_Gal17}
Alex {Kendall} and Yarin {Gal}.
\newblock {What Uncertainties Do We Need in Bayesian Deep Learning for Computer
  Vision?}
\newblock {\em arXiv e-prints}, page arXiv:1703.04977, March 2017.

\bibitem{KingmaBa14}
Diederik~P. {Kingma} and Jimmy {Ba}.
\newblock {Adam: A Method for Stochastic Optimization}.
\newblock {\em arXiv e-prints}, page arXiv:1412.6980, December 2014.

\bibitem{AbdallaEtal22}
Elcio {Abdalla}, Guillermo~Franco {Abell{\'a}n}, Amin {Aboubrahim}, Adriano
  {Agnello}, {\"O}zg{\"u}r {Akarsu}, Yashar {Akrami}, George {Alestas}, Daniel
  {Aloni}, Luca {Amendola}, Luis~A. {Anchordoqui}, Richard~I. {Anderson}, Nikki
  {Arendse}, Marika {Asgari}, Mario {Ballardini}, Vernon {Barger}, Spyros
  {Basilakos}, Ronaldo~C. {Batista}, Elia~S. {Battistelli}, Richard {Battye},
  Micol {Benetti}, David {Benisty}, Asher {Berlin}, Paolo {de Bernardis},
  Emanuele {Berti}, Bohdan {Bidenko}, Simon {Birrer}, John~P. {Blakeslee},
  Kimberly~K. {Boddy}, Clecio~R. {Bom}, Alexander {Bonilla}, Nicola {Borghi},
  Fran{\c{c}}ois~R. {Bouchet}, Matteo {Braglia}, Thomas {Buchert}, Elizabeth
  {Buckley-Geer}, Erminia {Calabrese}, Robert~R. {Caldwell}, David {Camarena},
  Salvatore {Capozziello}, Stefano {Casertano}, Geoff C.~F. {Chen}, Jens
  {Chluba}, Angela {Chen}, Hsin-Yu {Chen}, Anton {Chudaykin}, Michele {Cicoli},
  Craig~J. {Copi}, Fred {Courbin}, Francis-Yan {Cyr-Racine}, Bo{\.z}ena
  {Czerny}, Maria {Dainotti}, Guido {D'Amico}, Anne-Christine {Davis}, Javier
  {de Cruz P{\'e}rez}, Jaume {de Haro}, Jacques {Delabrouille}, Peter~B.
  {Denton}, Suhail {Dhawan}, Keith~R. {Dienes}, Eleonora {Di Valentino},
  Pu~{Du}, Dominique {Eckert}, Celia {Escamilla-Rivera}, Agn{\`e}s {Fert{\'e}},
  Fabio {Finelli}, Pablo {Fosalba}, Wendy~L. {Freedman}, Noemi {Frusciante},
  Enrique {Gazta{\~n}aga}, William {Giar{\`e}}, Elena {Giusarma}, Adri{\`a}
  {G{\'o}mez-Valent}, Will {Handley}, Ian {Harrison}, Luke {Hart}, Dhiraj~Kumar
  {Hazra}, Alan {Heavens}, Asta {Heinesen}, Hendrik {Hildebrandt}, J.~Colin
  {Hill}, Natalie~B. {Hogg}, Daniel~E. {Holz}, Deanna~C. {Hooper}, Nikoo
  {Hosseininejad}, Dragan {Huterer}, Mustapha {Ishak}, Mikhail~M. {Ivanov},
  Andrew~H. {Jaffe}, In~Sung {Jang}, Karsten {Jedamzik}, Raul {Jimenez},
  Melissa {Joseph}, Shahab {Joudaki}, Marc {Kamionkowski}, Tanvi {Karwal},
  Lavrentios {Kazantzidis}, Ryan~E. {Keeley}, Michael {Klasen}, Eiichiro
  {Komatsu}, L{\'e}on V.~E. {Koopmans}, Suresh {Kumar}, Luca {Lamagna}, Ruth
  {Lazkoz}, Chung-Chi {Lee}, Julien {Lesgourgues}, Jackson {Levi Said},
  Tiffany~R. {Lewis}, Benjamin {L'Huillier}, Matteo {Lucca}, Roy {Maartens},
  Lucas~M. {Macri}, Danny {Marfatia}, Valerio {Marra}, Carlos J.~A.~P.
  {Martins}, Silvia {Masi}, Sabino {Matarrese}, Arindam {Mazumdar}, Alessandro
  {Melchiorri}, Olga {Mena}, Laura {Mersini-Houghton}, James {Mertens}, Dinko
  {Milakovi{\'c}}, Yuto {Minami}, Vivian {Miranda}, Cristian {Moreno-Pulido},
  Michele {Moresco}, David~F. {Mota}, Emil {Mottola}, Simone {Mozzon}, Jessica
  {Muir}, Ankan {Mukherjee}, Suvodip {Mukherjee}, Pavel {Naselsky}, Pran
  {Nath}, Savvas {Nesseris}, Florian {Niedermann}, Alessio {Notari}, Rafael~C.
  {Nunes}, Eoin {{\'O} Colg{\'a}in}, Kayla~A. {Owens}, Emre {{\"O}z{\"u}lker},
  Francesco {Pace}, Andronikos {Paliathanasis}, Antonella {Palmese}, Supriya
  {Pan}, Daniela {Paoletti}, Santiago~E. {Perez Bergliaffa}, Leandros
  {Perivolaropoulos}, Dominic~W. {Pesce}, Valeria {Pettorino}, Oliver H.~E.
  {Philcox}, Levon {Pogosian}, Vivian {Poulin}, Gaspard {Poulot}, Marco
  {Raveri}, Mark~J. {Reid}, Fabrizio {Renzi}, Adam~G. {Riess}, Vivian~I.
  {Sabla}, Paolo {Salucci}, Vincenzo {Salzano}, Emmanuel~N. {Saridakis},
  Bangalore~S. {Sathyaprakash}, Martin {Schmaltz}, Nils {Sch{\"o}neberg}, Dan
  {Scolnic}, Anjan~A. {Sen}, Neelima {Sehgal}, Arman {Shafieloo}, M.~M.
  {Sheikh-Jabbari}, Joseph {Silk}, Alessandra {Silvestri}, Foteini {Skara},
  Martin~S. {Sloth}, Marcelle {Soares-Santos}, Joan {Sol{\`a} Peracaula},
  Yu-Yang {Songsheng}, Jorge~F. {Soriano}, Denitsa {Staicova}, Glenn~D.
  {Starkman}, Istv{\'a}n {Szapudi}, Elsa~M. {Teixeira}, Brooks {Thomas},
  Tommaso {Treu}, Emery {Trott}, Carsten {van de Bruck}, J.~Alberto {Vazquez},
  Licia {Verde}, Luca {Visinelli}, Deng {Wang}, Jian-Min {Wang}, Shao-Jiang
  {Wang}, Richard {Watkins}, Scott {Watson}, John~K. {Webb}, Neal {Weiner},
  Amanda {Weltman}, Samuel~J. {Witte}, Rados{\l}aw {Wojtak}, Anil~Kumar
  {Yadav}, Weiqiang {Yang}, Gong-Bo {Zhao}, and Miguel {Zumalac{\'a}rregui}.
\newblock {Cosmology intertwined: A review of the particle physics,
  astrophysics, and cosmology associated with the cosmological tensions and
  anomalies}.
\newblock {\em Journal of High Energy Astrophysics}, 34:49--211, June 2022.

\bibitem{suyuEtal12b}
S.~H. {Suyu}, T.~{Treu}, R.~D. {Blandford}, W.~L. {Freedman}, S.~{Hilbert},
  C.~{Blake}, J.~{Braatz}, F.~{Courbin}, J.~{Dunkley}, L.~{Greenhill},
  E.~{Humphreys}, S.~{Jha}, R.~{Kirshner}, K.~Y. {Lo}, L.~{Macri}, B.~F.
  {Madore}, P.~J. {Marshall}, G.~{Meylan}, J.~{Mould}, B.~{Reid}, M.~{Reid},
  A.~{Riess}, D.~{Schlegel}, V.~{Scowcroft}, and L.~{Verde}.
\newblock {The Hubble constant and new discoveries in cosmology}.
\newblock {\em ArXiv e-prints (1202.4459)}, February 2012.

\bibitem{FalcoEtal85}
E.~E. {Falco}, M.~V. {Gorenstein}, and I.~I. {Shapiro}.
\newblock {On model-dependent bounds on H(0) from gravitational images
  Application of Q0957 + 561A,B}.
\newblock {\em \apjl}, 289:L1--L4, February 1985.

\bibitem{GorensteinEtal88}
M.~V. {Gorenstein}, E.~E. {Falco}, and I.~I. {Shapiro}.
\newblock {Degeneracies in Parameter Estimates for Models of Gravitational Lens
  Systems}.
\newblock {\em \apj}, 327:693, April 1988.

\bibitem{SchneiderSluse13}
P.~{Schneider} and D.~{Sluse}.
\newblock {Mass-sheet degeneracy, power-law models and external convergence:
  Impact on the determination of the Hubble constant from gravitational
  lensing}.
\newblock {\em \aap}, 559:A37, November 2013.

\bibitem{XuEtal16}
D.~{Xu}, D.~{Sluse}, P.~{Schneider}, V.~{Springel}, M.~{Vogelsberger},
  D.~{Nelson}, and L.~{Hernquist}.
\newblock {Lens galaxies in the Illustris simulation: power-law models and the
  bias of the Hubble constant from time delays}.
\newblock {\em \mnras}, 456:739--755, February 2016.

\bibitem{GomerEtal20}
M.~{Gomer} and L.~L.~R. {Williams}.
\newblock {Galaxy-lens determination of H$_{0}$: constraining density slope in
  the context of the mass sheet degeneracy}.
\newblock {\em \jcap}, 2020(11):045, November 2020.

\bibitem{Kochanek20}
C.~S. {Kochanek}.
\newblock {Overconstrained gravitational lens models and the Hubble constant}.
\newblock {\em \mnras}, 493(2):1725--1735, April 2020.

\bibitem{BlumEtal20}
Kfir {Blum}, Emanuele {Castorina}, and Marko {Simonovi{\'c}}.
\newblock {Could Quasar Lensing Time Delays Hint to a Core Component in Halos,
  Instead of H$_{0}$ Tension?}
\newblock {\em \apjl}, 892(2):L27, April 2020.

\bibitem{MillonEtal20_TDCOSMOI}
M.~{Millon}, A.~{Galan}, F.~{Courbin}, T.~{Treu}, S.~H. {Suyu}, X.~{Ding},
  S.~{Birrer}, G.~C.~F. {Chen}, A.~J. {Shajib}, D.~{Sluse}, K.~C. {Wong},
  A.~{Agnello}, M.~W. {Auger}, E.~J. {Buckley-Geer}, J.~H.~H. {Chan},
  T.~{Collett}, C.~D. {Fassnacht}, S.~{Hilbert}, L.~V.~E. {Koopmans},
  V.~{Motta}, S.~{Mukherjee}, C.~E. {Rusu}, A.~{Sonnenfeld}, C.~{Spiniello},
  and L.~{Van de Vyvere}.
\newblock {TDCOSMO. I. An exploration of systematic uncertainties in the
  inference of H$_{0}$ from time-delay cosmography}.
\newblock {\em \aap}, 639:A101, July 2020.

\bibitem{DingEtal21}
X.~{Ding}, T.~{Treu}, S.~{Birrer}, G.~C.~F. {Chen}, J.~{Coles}, P.~{Denzel},
  M.~{Frigo}, A.~{Galan}, P.~J. {Marshall}, M.~{Millon}, A.~{More}, A.~J.
  {Shajib}, D.~{Sluse}, H.~{Tak}, D.~{Xu}, M.~W. {Auger}, V.~{Bonvin},
  H.~{Chand}, F.~{Courbin}, G.~{Despali}, C.~D. {Fassnacht}, D.~{Gilman},
  S.~{Hilbert}, S.~R. {Kumar}, J.~Y.~Y. {Lin}, J.~W. {Park}, P.~{Saha},
  S.~{Vegetti}, L.~{Van de Vyvere}, and L.~L.~R. {Williams}.
\newblock {Time delay lens modelling challenge}.
\newblock {\em \mnras}, 503(1):1096--1123, May 2021.

\bibitem{BirrerEtal20}
S.~{Birrer}, A.~J. {Shajib}, A.~{Galan}, M.~{Millon}, T.~{Treu}, A.~{Agnello},
  M.~{Auger}, G.~C.~F. {Chen}, L.~{Christensen}, T.~{Collett}, F.~{Courbin},
  C.~D. {Fassnacht}, L.~V.~E. {Koopmans}, P.~J. {Marshall}, J.~W. {Park}, C.~E.
  {Rusu}, D.~{Sluse}, C.~{Spiniello}, S.~H. {Suyu}, S.~{Wagner-Carena}, K.~C.
  {Wong}, M.~{Barnab{\`e}}, A.~S. {Bolton}, O.~{Czoske}, X.~{Ding}, J.~A.
  {Frieman}, and L.~{Van de Vyvere}.
\newblock {TDCOSMO. IV. Hierarchical time-delay cosmography {\textendash} joint
  inference of the Hubble constant and galaxy density profiles}.
\newblock {\em \aap}, 643:A165, November 2020.

\bibitem{GChenEtal21_TDCOSMOVI}
Geoff C.-F. {Chen}, Christopher~D. {Fassnacht}, Sherry~H. {Suyu}, Ak{\i}n
  {Y{\i}ld{\i}r{\i}m}, Eiichiro {Komatsu}, and Jos{\'e} {Luis Bernal}.
\newblock {TDCOSMO. VI. Distance measurements in time-delay cosmography under
  the mass-sheet transformation}.
\newblock {\em \aap}, 652:A7, August 2021.

\bibitem{TagoreEtal18}
Amitpal~S. {Tagore}, David~J. {Barnes}, Neal {Jackson}, Scott~T. {Kay},
  Matthieu {Schaller}, Joop {Schaye}, and Tom {Theuns}.
\newblock {Reducing biases on H$_{0}$ measurements using strong lensing and
  galaxy dynamics: results from the EAGLE simulation}.
\newblock {\em \mnras}, 474(3):3403--3422, March 2018.

\end{thebibliography}
\end{document}